\documentclass[12pt,psfig]{article}
\usepackage{amssymb}
\input{psfig}
\def\1E{1E 161348--5055}
\def\h{^{\rm h}}
\def\m{^{\rm m}}

\def\HII{H\,{\sc ii}}
\def\HI{\rm H\,{\sc i}}
\def\mjb{mJy beam$^{-1}$}
\def\jb{Jy beam$^{-1}$}
\def\k{km s$^{-1}$}
\def\pp{^{\prime\prime}}
\def\cm2{cm$^{-2}$}
\def\c3{cm$^{-3}$}
\def\farcs{\hbox{$.\!\!^{\prime\prime}$}}
\def\la{{\hbox{$\lesssim$}}}

\def\h{^{\rm h}}

\def\m{^{\rm m}}
\def\fs{\hbox{$.\!\!{}^{\rm s}$}}

\begin{document}
\title{Influence of the neutron star \1E~ in RCW 103 on the surrounding medium}
\author{E. M. Reynoso, $^{1,2,}${\footnote{Member of the Carrera del
Investigador Cient\'\i fico, CONICET, Argentina.}}
 A. J. Green, $^{1}$
 S. Johnston, $^{1}$ 
 W. M. Goss, $^{3}$
 G. M. Dubner, $^{2,*}$ \and
 E. B. Giacani $^{2,*}$
} 

\date{}
\maketitle

{\center
$^1$ School of Physics, University of Sydney, NSW 2006, Australia\\ereynoso@iafe.uba.ar\\[3mm]
$^2$ Instituto de Astronom\'\i a y F\'\i sica del Espacio, CC 67,
Suc 28, 1428 Buenos Aires, Argentina\\[3mm]
$^3$ National Radio Astronomy Observatory, P. O. Box 0, Socorro, New Mexico 87801, USA\\[3mm]
}

\begin{abstract}
We have carried out a study of the neutral hydrogen in the direction of
the X-ray source \1E, a compact central object (CCO) located in the interior 
of the supernova remnant (SNR) RCW 103. The \HI~ 21 cm line observations were
carried out using the Australia Telescope Compact Array, complemented with
single dish data from the Parkes radio telescope to recover information at
all spatial scales. We derive a distance to RCW 103 of 3.3 kpc, in
agreement with previous distance measurements. We have also detected a 
small hole in the \HI~ emission which is positionally and kinematically
coincident with the location of the CCO which confirms the association
between the SNR and the CCO. This is the third case of a depression in 
\HI~ emission seemingly associated with CCOs in SNRs. The characteristic
parameters of the holes such as their size, eccentricity and evacuated
mass are similar in all three cases.  We estimate the absorbing \HI~ column
density towards \1E~ to be $\sim 6\times 10^{21}$ \cm2, a value
compatible with a blackbody solution for the CCO X-ray emission. However,
the implied brightness temperature is very high compared to most neutron
stars.  Moreover, the strong long-term variability in X-rays favours the
hypothesis that \1E~ is an accreting binary source rather than an
isolated, cooling neutron star. An analysis of the continuum image
obtained at 1.4 GHz from these observations shows no trace of a pulsar
wind nebula around \1E, in spite of it being a young object.
\end{abstract}

{\bf Keywords:}
stars: neutron -- supernova remnants -- ISM: individual: RCW 103 --
X--rays: individual: \1E~ -- spectral lines: neutral hydrogen.

\section{Introduction}

X-ray observations from the past few years have revealed the existence of
a large variety of Galactic point-like sources, many of them identified with neutron
stars but presenting very different observational properties. These unresolved
sources, which appear either isolated or in the interior of supernova remnants
(SNR), have no radio counterpart and very high X-ray to optical flux ratios. About
half of these sources show X-ray pulsations with periods between 6 and 12
seconds (anomalous X-ray pulsars, AXP) and can even present sporadic strong
$\gamma$-ray emission (soft gamma-ray repeaters, SGR). The rest of these
sources are generally called either isolated neutron stars (INS) if they are not
associated with any other object in the sky, or central compact objects (CCO)
if they appear projected on the interior of a SNR (see a recent review by
Pavlov et al. 2002).

The nature of most CCOs is still unclear. Brazier \& Johnston (1999) explain
them as normal pulsars with unfavourable radio beaming,  while the soft thermal
X-rays would be easily detected due to its almost isotropic emission. On the
other hand, Vasisht et al. (1997) and Frail (1998) conclude that CCOs are
neutron stars (NSs) born with long initial periods and high magnetic fields
($B>10^{14}$ G), and thus would be related to AXPs and SGRs. However, Geppert,
Page \& Zannias (1999) suggest that CCOs are fast-spinning, weakly-magnetized
sources. The X-ray emission from CCOs is generally explained as thermal
radiation from cooling NSs (e.g Zavlin, Tr\"umper \& Pavlov 1999), with
typical temperatures of a few 10$^6$ K, as inferred from their thermal-like
spectra.

In a recent \HI \ study towards the bilateral
SNR G296.5+10.0, Giacani et al. (2000) found that the associated CCO, 1E
1207.4--5209, lies near the center of a small \HI \ depression located at the same
systemic velocity as the SNR. The authors propose that the depression is due
to self-absorption of a cool \HI \ cloud just foreground to a hotter volume of
gas surrounding the CCO, and heated by its X-ray flux. We have begun a
systematic search for similar traces in the neutral gas around other CCOs. The
observations towards the X-ray point source RX J0822--4300 in Puppis A revealed
an \HI \ structure consisting of a nearly circular minimum coincident with the CCO
plus two aligned lobe-like depressions that appear to emerge from the CCO (Reynoso 
et al. 2003).
The two lobes appear to have been formed by a combination of the proper motion
of the CCO and the ejection of a collimated outflow. In this paper, we present
the results obtained for \1E, the CCO associated with the SNR RCW
103 (G332.4--0.4).

At radio wavelengths, RCW 103 appears as an almost complete, circular
$8^\prime$ diameter shell (Caswell et al. 1980). High-resolution radio
polarimetric data reveal an approximately east-west alignment of the magnetic
field which could be helping to limit the expansion north-south (Dickel et
al. 1996). The brightening of the rim in the northern and southern sides,
that hints for an incipient bilateral barrel shape for this SNR, could be
related to this frozen-in magnetic field.
Based on \HI \ absorption measurements at 21 cm, Caswell et al. (1975)
suggest a distance of 3.3 kpc. Optical filaments are seen
toward the brighter regions of the radio shell (van den Bergh, Marscher \&
Terzian 1973; Ruiz 1983), and observations at infrared wavelengths show
evidence for interaction with a dense interstellar medium (ISM),
particularly to the south (Oliva, Moorwood \& Danziger 1990; Burton \&
Spyromilio 1993; Oliva et al. 1999). An optical expansion study of RCW 103
(Carter,
Dickel \& Bomans 1997) indicates that this SNR is probably about 2,000 years
old, based on an assumed distance of 3.3 kpc. However, optical extinction studies
suggest distances around 6.5 kpc (Leibowitz \& Danziger 1983; Ruiz 1983).

Soft X-ray emission from RCW 103 was first detected with the Einstein
Observatory (Tuohy et al. 1979), indicating a close correlation with the 
non-thermal radio emission. Also, a faint, point X-ray source, \1E ,
was located near the centre of the SNR (Tuohy \& Garmire 1980). The lack of
optical or radio counterparts led Tuohy et al. (1983) to propose
that this point source was a thermally radiating CCO. However, subsequent
observations with the Einstein IPC and ROSAT failed to confirm the
existence of this source (Becker et al. 1993). Finally, Gotthelf, Petre
\& Hwang (1997) detected hard X-rays from the elusive CCO
using ASCA. They found that the spectral characteristics were incompatible
with a simple cooling NS model. Follow-up observations (Gotthelf, Petre
\& Vasisht 1999) confirmed that \1E~ manifests long-term variability,
explaining the non-detections after its discovery. There are strong
indications that \1E~ is part of a binary system (see \S4.1).

In this paper, we present radio contiuum at 1.38 GHz and \HI \ 
$\lambda$21 cm observations carried out with the Australia Telescope 
Compact Array (ATCA) towards the SNR RCW 103.

\section{Observations and data reduction}

Interferometric observations were obtained with the ATCA during one
session of 12 h with the 750A array (baselines from 76.5 to 735 m), on
2002 January 22, and one session of 12 h with the EW 367 array (baselines
from 46 to 367 m) on 2002 April 1. The antennas were pointed at RA$=16\h
17\m 30\fs 0$, Dec.$= -51^\circ 0^\prime 0\pp$ (J2000).  A correlator
configuration of 1024 channels, covering a total bandwith of 4 MHz centered
at 1420 MHz, was used. The corresponding velocity resolution at this
frequency is 1 \k. Continuum data were obtained simultaneously with a
bandwidth of 128 MHz centered at 1384 MHz. The source PKS B1934--638 was
used for primary flux density calibration, while phases were calibrated
with PKS B1657--56.

The data were processed with the MIRIAD software package (Sault, Tauben \&
Wright 1995). To subtract the continuum component from the \HI \ data set,
a linear baseline was fitted to 370 line-free channels. The final \HI \ cube
was constructed with a 15$\pp$ cell size,
covering an area of $64^\prime \times 64^\prime$ and keeping 350 channels from
--150 to +138 \k. Sidelobes were supressed over an area of $34^\prime \times
34^\prime$. The image was cleaned and convolved with a 50$\pp$ Gaussian beam. 
The continuum image was constructed with the same geometry and angular resolution 
as the \HI \ data. After cleaning and restoring with a 50$\pp$ Gaussian beam, a 
sensitivity of $\sigma \simeq 5.5$ \mjb \ was achieved.

To recover structures at shorter spatial frequencies, the ATCA \HI \ data
were combined in the {\it u,v} plane with single
dish data from the Southern Galactic Plane Survey (SGPS; McClure--Griffiths
et al. 2001) obtained with the Parkes telescope. In order to match the
intensity units in both sets, a factor of 257.5 was applied to the ATCA
data cube to convert flux densities to brightness temperatures. No tapering
was applied to the low resolution cube. The rms of the combined image is 2 K
per channel.

\begin{figure}
\caption{Radio continuum image centred at RCW 103, obtained with the
ATCA at 1384 MHz. The gray scale is shown on top of the image in units
of \jb. The beam, $50\pp \times 50\pp$, is plotted in the bottom
right corner. The noise level is 5.5 \mjb.}
\label{fig1}
\end{figure}

\section {Results}

\subsection{Absorption study}

\begin{figure}
\caption{(a) Average \HI \ emission profile towards RCW 103 using interferometric
and single dish data. (b) Absorption profile towards RCW 103 based on
interferometric ATCA data. (c) Same as (b) but excluding baselines shorter
than 1 k$\lambda$. In all cases, the velocity resolution is 1 \k .}
\label{fig2}
\end{figure}

To constrain the systemic velocity of RCW 103, we analyzed the \HI \
absorption against the strong radio continuum emission from the remnant.
The radio contimuum image obtained with ATCA at the frequency of 1384
MHz in a field of about $45^\prime$ around RCW 103 is shown in Figure
1. According to the equation of radiative transfer, the emission $T_{b_v}$
at a velocity $v$ measured by an interferometer is given by
\begin{equation}T_{b_v} = T_s (1 - e^{-\tau_v} ) + T_c e^{-\tau_v},\end{equation}
where $T_s$ and $\tau_v$ are the spin temperature and the optical depth of
the intervening \HI \ gas, and $T_c$ is the background continuum
emission. Since we have subtracted the radio continuum from out \HI \ data,
then eq.(1) can be reduced to
\begin{equation}T_{L_v} = (T_s - T_c) (1 - e^{-\tau_v}),\end{equation} where
$T_{L_v} = T_{b_v} - T_c$. Equation (1) can be simplified to a single unknown
if $T_c \gg T_s$.  This condition can be accomplished by using only
interferometric data, since in such way the extended \HI \ emission is
filtered out.

Figure 2a shows an average \HI \ emission profile towards RCW 103 combining
interferometric plus single dish data, while Fig. 2b shows an average
profile of $e^{-\tau_v}$ over the same area, computed as described above
(eq. (2) ). Strong absorption peaks appear at --43, --17, +1, and +34
\k . All but the last feature were detected by Caswell et al. (1975)
based on data obtained with the Parkes interferometer with a 2 \k \ spectral
resolution. We do not believe the absorption at +34 \k \ to be real because
(a) absorption features should appear also from --43 \k \ to the tangent point
at --115 \k, and (b) the distance to the SNR would be $\sim 20$ kpc, which
gives an unrealistic size and expansion velocity.

For comparison, we obtained ATCA spectra towards the
strong Galactic radio sources present in the field G332.7--0.6
and G332.2--0.4 (Fig. 1), placed within $15^\prime$ of RCW
103, and found that absorption features appear only in the
velocity ranges from 0 to --50 and to --58 \k \ respectively.

To solve this puzzle, we followed the method employed by Dickey
et al. (2003). We constructed a new \HI \ cube without subtracting
the radio continuum and removing all baselines shorter than 1
k$\lambda$ (210 m). In this way, all \HI \ structures larger
than $\sim 4^\prime$ are filtered out. The term involving $T_s$ in
eq. (1) can thus be neglected. The data cube was further weighted
by the radio continuum. We then computed $e^{-\tau_v}$ using eq.
(1). The resulting profile is plotted in Figure 2c.

This filtering method succeeded in removing most of the \HI \
emission, as is apparent by comparing Figs. 2b and 2c. The absorption
feature at --43 \k \ is confirmed, while the feature at +34 \k has
disappeared. This implies that the latter was produced by self
absorption on a scale between $4^\prime$ and $\sim 30^\prime$.

The absorption feature at --43 \k \ corresponds to a lower distance
limit of 3.1 kpc according to the Galactic rotation model of Fich, Blitz
\& Stark (1989).  We do not see any absorption against the emission peak
at $\sim -75$ \k, even though it has a brightness temperature of $\sim
80$ K. Thus the upper distance limit is 4.6 kpc. The lower distance
coincides with the near side of the Scutum-Crux arm (Georgelin \& Georgelin
1976), and will be assumed throughout the paper.

\begin{figure}
\caption{Gray-scale and contour image of the average \HI \
emission between --46.1 and --53.5 \k \ towards RCW 103. The rms is 2K.
The beam, $50\pp \times 50\pp$, is plotted as a white open circle in the
bottom left corner. The brightness-temperature scale is indicated on top
of the image in units of K, while the contour levels vary from 68 K
in steps of 6 K. For comparison, a few representative contours of the
radio continuum emission are included as white lines. The cross
indicates the position of \1E as given by Garmire et al. (2000a). }
\label{fig3}
\end{figure}

\subsection{\HI \ associated with RCW 103}

We estimated
the \HI \ column density towards \1E~ by integrating the foreground
\HI \ brightness temperature of the interferometric plus single-dish
data up to --43 \k . The value obtained depends on the lower limit
adopted for the velocity interval in the integration since, although
the local gas is supposed to lie at a systemic velocity of 0 \k,
turbulence may cause departures of about 7 \k \ from this value. Besides, 
due to the distance ambiguity, it is possible that some background gas is 
included in the integration, leading to an overestimation of $N_{\rm H}$. 
At $v\simeq -10$ \k, the line of sight crosses the far side of the
Scutum-Crux arm. Taking into account these considerations, we integrate
the brightness temperature between $\sim +3.5$ and $-43$ \k \ and
estimate the \HI \ column density to be $N_{\rm H} \la 6 \times
10^{21}$ \cm2. The implicances of this determination will be discussed
in $\S$4.

An inspection of the \HI \ images at velocities around \- --43\- \k \
revealed that \1E~ lies inside a local \HI \ depression which is present
in all channels between --46.1 and --53.5 \k . In Figure 3, an image of 
the average \HI \ emission within this velocity interval is shown. The 
\HI \ depression, which attains its minimum at RA$=16\h 17\m 34\fs 8$, 
Dec.$= -51^\circ 2^\prime 40\pp$ (J2000), is elongated, with a 
minor-to-major axis ratio of $\sim 0.7$, and has a mean diameter of 
$\sim 64\pp$ (1 pc at a distance of 3.1 kpc). \1E~ is $20\pp$ (0.3 pc) 
away from the centre of the hole. We note that due to the proximity
to the Galactic Plane, our data show several similar \HI \ depressions 
at different locations and velocities. However, the coincidence in
position and velocity makes an association between the CCO and this
\HI \ feature very likely. 

In what follows, we analyze possible origins for this \HI \ void.
If the \HI \ minimum is produced because of a real absence of \HI, then
the missing mass is estimated to be 0.3 M$_\odot$. If instead it is
produced by hot \HI \ gas self-absorbed by a cooler foreground, as
proposed for the CCO 1E 1207.4--5209 (Giacani et al. 2000), then we
can follow the method described by Schwarz et al. (1995) to estimate
an upper limit for the temperature of the hot neutral hydrogen gas.

If self-absorption is considered, then eq. (1) can be re-written as
(cf. eq. (3) in Schwarz et al. 1995)
\begin{equation}T_{L_v} = (T_s - T_{bg}) A_v - T_c A_v, \end{equation}
\noindent with $A_v = (1 - e^{-{\tau_v}})$, and where $T_{bg}$ is the
temperature of the background hot \HI. Assuming that $A_v$ is uniform
across the continuum source, then $T_{L_v}$ is a linear function of $T_c$
with slope $(-A_v)$ and a zero offset $(T_s - T_{bg}) A_v$. The \HI \
hole attains its minimum emission at $v=-46.1$ \k. We compared the
brightness temperature of the ATCA \HI \ data at this velocity with
the continuum emission and fitted a straight line to the distribution.
To avoid confusion possibly introduced by regions of low continuum
emission, all values of $T_c$ under 350 \mjb~ were clipped. We obtained
that $A_v \simeq 0.195$, which implies that $\tau_v\simeq 0.2$.
To estimate the spin temperature, it must be noticed that for a single
dish measurement towards a region with no continuum emission, the
brightness temperature of the line is $T_{L_v} = T_s A_v$ (Schwarz et al.
1995). We averaged the Parkes data in a box around RCW 103 (excluding
the contribution from the SNR) and obtained that $<\!\!T_{L_v}\!\!>\ \leq
100$ K, hence $T_s \leq 20$ K. Finally, at the location of \1E, the
interferometric data give $T_c=44$ K and $T_{L_v} = -30$ K.  Thus, applying
eq. (3) to the emission towards the CCO, the hot \HI \ gas in the hole
is found to have a temperature of $\leq 130$ K. Such a temperature is
unrealistically low at the interior of a SNR, therefore we discard the
possibility that the \HI \ minima observed around CCOs are produced by
self-absorption.

\begin{table*}
 \centering
 \begin{minipage}{140mm}
  \caption{Parameters of the \HI \ depressions around CCOs}
  \begin{tabular}{@{}lccc@{}}
  \hline
   CCO& 1E 1207.4--5209\footnote{Giacani et al. (2000)}& RX J0822--4300
\footnote{Reynoso et al. (2003)}&\1E \footnote{Present work.}\\
   (associated SNR)& (G296.5+10.0) & (Puppis A) & (RCW 103)\\
 \hline
mean angular diameter ($^\prime$)& 5.3 & 2.3 & 1.1 \\
mean linear diameter (pc)& 3.2 & 1.5 & 1.0 \\
minor/major axis ratio & 0.8 & 0.8 & 0.7 \\
CCO offset from centre ($\pp$)& 30 & 37 & 20 \\
CCO offset from centre (pc) & 0.3 & 0.4 & 0.3 \\
missing mass (M$_\odot$) & -- & 0.1 & 0.3 \\
\hline
\end{tabular}
\end{minipage}
\end{table*}

\subsection {Radio continuum emission}

The rotational energy of pulsars is 
dissipated  via a magnetized relativistic wind composed of electrons 
and positrons. The shock front between this wind and the ambient 
medium can give rise to a synchrotron emitting bubble (known as a 
pulsar wind nebula, PWN). The detectability of such a PWN depends 
on the density of the ambient medium and the pulsar parameters. The 
majority of pulsars do not have detectable PWN (Gaensler et al. 2000). 
For a very young pulsar inside an SNR, the luminosity $L$ of the PWN 
is a strong function of the initial period of the pulsar, $P_{0}$ 
($L \propto P_{0}^{-5}$; Reynolds and Chevalier 1984).

We do not detect a PWN at the position of \1E. Rather, the radio 
continuum emission in RCW 103 has a minimum near the CCO (Figure 
4). To investigate if there is a PWN around the CCO that could be 
beam-diluted at the resolution of $50\pp$, we constructed another 
continuum image using only the longest baselines. This image, with 
a resolution of $6\farcs 2 \times 4\farcs 6$, shows no emission at 
the position of the CCO down to a level of $5\sigma = 1$ \mjb. 
This limit is not very constraining - it may imply either that the 
spin period of \1E \ is not particularly short or that the PWN is 
not well confined in the interior of the SNR.

\begin{figure}
\caption{Gray-scale and contour image of RCW 103 at the
radio continuum frequency of 1384 MHz. The gray scale is shown on top
of the image in units of \jb. The contours are plotted in steps
of 7.5\%  of the peak intensity, 768 \mjb, starting at 15\%. For
clarity purposes, white lines are used over dark background. The 
cross shows the position of \1E as given by Garmire et al. (2000a).
The noise level is 5.5 \mjb.}
\label{fig4}
\end{figure}

\section {Discussion}

\subsection{Spectral model}

Gotthelf et al. (1997) investigated different fits to the X-ray spectrum
of \1E using
a nonequilibrium ionization plasma model combined alternatively with
a blackbody, a power-law, and a thermal bremsstrahlung continuum component.
The best fits were achieved with \HI \ column densities of 5, 31, and $16
\times 10^{21}$ \cm2 \ respectively, although a fixed value of $N_{\rm H}=6.7
\times 10^{21}$ \cm2 \ also gives reasonable fits in all three cases. The
value of $N_{\rm H}$ that we estimated from the present observations,
$N_{\rm H}\leq 6 \times 10^{21}$ \cm2, clearly favours the blackbody
solution in this case.

However, since 1997, there is growing evidence to indicate that \1E 
might be an accreting binary system. Chandra and ASCA observations 
revealed a sinusoidal light curve with a period of $\sim 6.4$ hr 
(Garmire et al. 2000a). Even though subsequent observations failed 
to detect the modulation again (Garmire et al. 2000b) due to pile-up 
in the detectors from the very high X-ray flux, the detection of two 
partial dips separated by 180$^\circ$ in phase in the light curve, 
and a possible near-IR counterpart about $2\pp$ away from the nominal 
Chandra position, suggest that \1E~ may be powered by accretion likely 
from a low-mass companion in a binary system (Sanwal et al. 2002).
Finally, Becker \& Aschenbach (2002) discovered an eclipse 3 h after 
the start of their observations using XMM-Newton data. This would make 
\1E the first accreting binary detected inside a SNR. Becker \& 
Aschenbach (2002) also obtained an X-ray spectrum towards \1E \ and 
found that a simple blackbody model describes the data almost up to 
$\sim 5$ keV but not beyond. They obtain a best fit with a double 
blackbody, but the required $N_{\rm H}$ of $\leq 18 \times 10^{21}$
\cm2 \ is incompatible with our results.

We note that all the
blackbody models imply temperatures as high as $8-10 \times 10^6$ K
(Becker \& Aschenbach 2002) and a luminosity of $\sim 10^{34}$ ergs s$^{-1}$
(Gotthelf et al. 1997), at least twice the values predicted by theoretical
models (see Gotthelf et al. 1997 and references therein). In general, fitting
NS models to the compact sources inside SNRs presents the problem that the
derived parameters (temperature and luminosity) are far larger than normally
expected (Lloyd, Hernquist \& Heyl 2002). In addition, estimates of X-rays 
emitting areas based on blackbody models are usually too small to be the size 
of a NS (e.g. Becker \& Aschenbach 2002). One possible solution is proposed by
Heyl \& Hernquist (1998), who show that a hot blackbody can be mimicked if the
cooling NS model includes an ultramagnetized star (B $\sim 10^{15}$ G) with 
an accreted hydrogen atmosphere. This explanation was also invoked for
other CCOs, like RX J0822--43 in Puppis A (Zavlin et al. 1999) and 1E
1207.4--5209 in G296.5+10.0 (Zavlin et al. 1998). Even in non-magnetic
atmospheres, the effect of H or He atmospheres is to deviate the
high energy spectrum in such a way that the effective temperatures as
given by blackbody models could be overestimated by a factor of 2 (Zavlin, 
Pavlov \& Shibanov 1996). 

In summary, it seems most likely that \1E~ is a binary system and that 
fits to the X-ray spectrum need to be re-visited. On the other hand, it 
is still possible that this system is a single neutron star. The constraints 
we have put on $N_{\rm H}$ show that it is very hot for its age, and may 
indicate exotic atmospheric processes at work.

\subsection {The {\HI} depression}

The present study reveals the third case in which a CCO associated
with a SNR is located inside an \HI \ depression. The other two cases are 1E
1207.4--5209 in the SNR  G296.5+10.0, and RX J0822--4300 in Puppis A. In
Table 1 we compare the main parameters derived for the three
\HI \ depressions. In G296.5+10.0, only the size and velocity width are 
given in Giacani et al. (2000), thus the remaining parameters, where 
possible, have been estimated directly on the images presented in their 
paper. In computing the minor/major axis ratio, the beam elongation was 
taken into account.

A number of similarities can be found between the three cases. All
\HI \ cavities appear to have the same elongation, and the CCOs
are off-centred by similar distances. In the two cases where the
missing mass was computed, a similar value was obtained. On the other
hand, the size of the \HI \ hole in G296.5+10.0 is larger than
the other two holes by more than a factor of 2. This discrepancy
may be related to the beam size of the observations, which is
almost 4 times larger than those of the Puppis A and RCW 103
data. The main difference between the three cases is given by Puppis
A, which presents two lobe-like \HI \ depressions emerging from its
associated CCO at the systemic velocity of the SNR (Reynoso et al.
2003). The present observations do not reveal any similar morphology
around \1E.

In what follows we will analyze if the \HI \ depression around \1E
can be a swept up hole. The rate of energy needed to set the missing
mass in the cavity into motion with velocity $V=r/t$ is
\begin{equation}\dot E_k = 3 \times 10^{47} M r^2 t^{-3}{\rm \ erg\
s}^{-1}, \end{equation}
\noindent where $M$ is the evacuated mass in M$_\odot$, $r$ is the
radius of the hole in pc, and $t$ is the age of the CCO, equal to
the age of the host SNR, in years. Replacing in eq. (4) the radius and
missing mass computed in \S3, and assuming that the age of RCW 103
is 2,000 yrs (Carter et al. 1997), we obtain $\dot E_k = 3{^{+20}_{-2}}
\times 10^{36}$ erg s$^{-1}$, where the quoted errors allow for ages of
1,000 and 3,000 yrs old. The spin down energy loss has not been measured
for \1E, but if it is similar to the values observed in other CCOs 
(energy loss rates between $\sim 1 \times 10^{36}$ erg and $1.5 \times 
10^{37}$ erg; Slane et al. 1997, Brazier \& Johnston 1999), then a 
complete conversion of the rotational energy into kinetic energy of
the surrounding medium can account for the observed \HI \ hole.

An alternative possibility is that the depression does not contain swept
up, low density gas, but is filled with \HI \ gas heated up by the CCO at
temperatures higher than the surroundings. In that case, the depression
should contain enhanced ionized hydrogen (i.e. it would form a small \HII~ 
region), and this is not observed in the radiocontinuum image. We would, 
however, expect to see infra-red emission from such a region but a search 
in mid- and near-infrared wavelengths using data from the Midcourse Space 
Experiment (MSX) and the Two Micron All Sky Survey (2MASS), yielded negative 
results. In conclusion, we find that the swept-up cavity provides a more 
convincing explanation for the observed \HI \ minimum.

\section {Conclusions}

In this paper, we present the third case in which a CCO lies at a local
\HI~ minimum at a velocity compatible with the systemic velocity of the
host SNR. We have shown that self-absorption does not provide a
satisfactory explanation  for this kind of features, as was proposed
for 1E 1207.4--5209 (Giacani et al. 2000). Instead, it is possible that 
the \HI~ depression is a swept up hole, where $\sim 0.3 M_\odot$ of 
neutral gas has been evacuated. We have found a number of similarities
between the three \HI~ holes around CCOs detected so far, like the
elongation, the missing mass, and the off-centred position of the CCOs.
We did not detect any synchrotron nebula around the CCO down to a level
of 1 \mjb.

The present data allow us to constrain the \HI~ column density to be 
$N_H \sim 6\times 10^{21}$ \cm2. This column density favours the 
blackbody model of Gotthelf et al. (1999), however the derived 
temperature is too high to be explained by standard NS cooling models. 
Instead, it appears most likely that \1E~ is an accreting binary, and 
may constitute the first case in which such a system occurs in the 
interior of a SNR.

\section*{Acknowledgments}

We acknowledge the invaluable help of Naomi McClure-Griffiths, who
provided us the data from the SGPS and assisted us in combining them
with the ATCA observations. EMR acknowledges a Postdoctoral External
fellowship from CONICET, Argentina. This publication makes use of data
products from the Two Micron All Sky Survey, which is a joint project
of the University of Massachusetts and the Infrared Processing and
Analysis Center, funded by the National Aeronautics and Space
Administration and the National Science Foundation. The MOST is
operated by the University of Sydney with support from the Australian
Research Council and the Science Foundation for Physics within the
University of Sydney. This research was partially funded through
CONICET grant 4203/96 and grant UBACYT A013. The National Radio Astronomy
Observatory is a facility of the National Science Foundation operated
under a cooperative agreement by Associated Universities, Inc. The
Australia Telescope is funded by the Commonwealth of Australia for
operation as a National Facility managed by the CSIRO.

\section*{References}
Becker W., Aschenbach, B., 2002, Proc. of the 270 WE-Heraerus Seminar on Neutron 
Stars, Pulsars and Supernova Remnants. Eds. W. Becker, H. Lesch, and J. Tr\"umper,
Garching bei M\"unchen, MPE, p 64\\
Becker W., Tr\"umper J., Hasinger J., Aschenbach B., 1993, in Isolated Pulsars, 
edited by K. A. van Riper, R. I. Epstein, and C.  Ho (Cambridge University Press, 
Cambridge), p116\\
Brazier K. T.  S., Johnston S., 1999, MNRAS, 305, 671\\
Burton M. G., Spyromilio J., 1993, PASA, 10, 327\\
Carter L. M., Dickel J. R., Bomans D. J., 1997, PASA, 109, 990\\
Caswell J. L., Murray J. D., Roger R. S., Cole D. J., Cook D. J., 1975, A\&A, 
45, 239\\
Caswell J. L., Haynes R. F., Milne D. K., Wellington K., 1980, MNRAS, 190, 881\\
Dickel J. R., Green A., Ye T., Milne D. K., 1996, AJ, 111,340\\
Dickey J. M., McClure-Griffiths N. M., Gaensler B. M., Green A. J., 2003, ApJ, 
585, 801\\
Fich M., Blitz L., Stark A. A., 1989, ApJ, 342, 272\\
Frail D. A., 1998, in The Many Faces of Neutron Stars, ed. R. Buccheri, J. van 
Paradijs, \& M. A. Alpar (NATO ASI Ser. C, 515; Dordrecht: Kluwer), 179\\
Gaensler B. M., Stappers B. W., Frail D. A., Moffett D. A., Johnston S., 
Chatterjee S., 2000, MNRAS, 318, 58\\
Garmire G. P., Pavlov G. G., Garmire A. B., Zavlin V.  E., 2000a, IAU Circ. 7350\\
Garmire G. P., Garmire A. B., Pavlov G. G., Burrows, D. N., 2000b, AAS HEAD 
Meeting 32, \#32.11\\
Georgelin Y. M., Georgelin Y. P., 1976, A\&A, 49, 57\\
Geppert U., Page D., Zannias T., 1999, A\&A, 345, 847\\
Giacani E. B., Dubner G. M., Green A. J., Goss W. M., Gaensler B. M., 2000, AJ, 
119, 281\\
Gotthelf E. V., Petre R., Hwang U., 1997, ApJ, 487, L175\\
Gotthelf E. V., Petre R., Vasisht G., 1999, ApJ, 514, L107\\
Heyl J. S., Hernquist L., 1998, MNRAS, 298, 17\\
Leibowitz E. M., Danziger I. J., 1993, MNRAS, 204, 273\\
Lloyd, D. A., Hernquist, L., Heyl, J. S., 2002, in ASP Conf. Ser. 271, 
Constraining the Birth Events of Neutron Stars, eds. P. O. Slane and B. M. 
Gaensler (San Francisco: ASP), 323\\
McClure--Griffiths N. M., Green A. J., Dickey J. M., Gaensler B. M., Haynes
R. F., Wieringa M. H. 2001, ApJ, 551, 394\\
Oliva E., Moorwood A. F. M., Danziger I. J., 1990, A\&A, 240, 453\\
Oliva E., Moorwood A. F. M., Drapatz S., Lutz D., Strum E., 1999, A\&A, 343, 
9430\\
Pavlov G. G., Zavlin V. E., Tr\"umper J., 1999, ApJ, 511, L45\\
Pavlov, G.G., Sanwal, D., Garmire, G. P., Zavlin, V. E., 2002, ``The puzzling 
compact objects  in supernova remnants'', in Neutron Stars and Supernova 
Remnants, eds. O. Slane and B. M. Gaensler, ASP Conf. Ser., V. 2171, p247 
(astro-ph/0112322)\\
Reynolds, S. P., Chevalier, R. S., 1984, ApJ, 278, 630\\
Reynoso E. M., Green A. J., Johnston S., Dubner G. M., Giacani E. B., Goss W. 
M., 2003, MNRAS, 345, 671\\
Ruiz M. T., 1983, AJ, 88, 1210\\
Sanwal D., Garmire G. P., Garmire A., Pavlov G. G., Mignani R., 2002, BAAS, 
200, 7201\\
Sault R. J., Teuben P. J., Wright M. C. H., 1995, in ASP Conf. Ser. 77, 
Astronomical Data Analysis Software and Systems IV, ed. R. A. Shaw, H. E. 
Payne, \& J. J. E.  Hayes (San Francisco: ASP), 433\\
Schwarz U. J., Goss W. M., Kalberla P. M., Benaglia P.,  1995, A\&A, 299, 193\\
Slane P., Seward F.  D., Bandiera R., Torii K., Tsunemi H., 1997, ApJ, 485, 221\\
Tuohy I. R., C\'ordova F. A., Garmire G. P., Mason K. O., Charles P. A., Walter 
F. M., Clark D. H., 1979, ApJ, 230, 37\\
Tuohy I. R., Garmire G. P., 1980, ApJ, 239, L107\\
Tuohy I. R., Garmire G. P., Manchester R. N., Dopita M. A., 1983, ApJ, 268, 778\\
van den Bergh S., Marscher A., Terzian Y., 1973, ApJS, 26, 19\\
Vasisht G., Kulkarni S. R., Anderson S. B., Hamilton T.  T., Kawai N., 1997, 
ApJ, 476, L43\\
Zavlin V. E., Pavlov G. G., Shibanov Yu. A., 1996, A\&A, 315, 141\\
Zavlin V. E., Pavlov G. G., Tr\"umper J., 1998, A\&A, 331, 821\\
Zavlin V. E., Tr\"umper J., Pavlov G. G., 1999, ApJ, 525, 959\\

\end{document}